\documentclass{elsart}
\usepackage[dvips]{graphicx}
\usepackage{comment}

\def\nuc#1#2{\relax\ifmmode{}^{#1}{\protect\text{#2}}\else${}^{#1}$#2\fi}

\begin{document}
\begin{frontmatter}

\bigskip

\title{Long range effects on the optical model of $^{6}$He around the Coulomb barrier.}

\author[famn,cna]{J. P. Fern\'andez-Garc\'{\i}a}, 
\author[famn,csic]{M. Rodr\'{\i}guez-Gallardo},
\author[famn,cna]{M. A. G. Alvarez}, 
\author[famn]{A. M. Moro}

\address[famn]{Depto. FAMN, Universidad de Sevilla, Apartado 1065, 41080 Sevilla, Spain.}
\address[cna]{Centro Nacional de Aceleradores, Universidad de Sevilla, 41092 Sevilla, Spain.}
\address[csic]{Instituto de Estructura de la Materia, CSIC, Serrano 123, 28006 Madrid, Spain.}
\begin{abstract}
\noindent
We present an optical model (OM) analysis of the elastic scattering data of the reactions $^6$He+\nuc{27}{Al} and $^6$He+\nuc{208}{Pb} at incident energies around the Coulomb barrier.
The bare part of the optical potential is constructed microscopically by means of a double folding procedure,  using the S\~ao Paulo prescription without any renormalization. This bare interaction is supplemented with a Coulomb dipole polarization (CDP) potential, which takes into account the effect of the dipole Coulomb interaction. For this CDP potential, we use an analytical formula derived from the semiclassical theory of Coulomb excitation. The rest of the optical potential is parametrized in terms of Woods-Saxon shapes.
In the \nuc{6}{He}+\nuc{208}{Pb} case, the analysis confirms the presence of long range components, in agreement with previous works.
Four-body Continuum-Discretized Coupled-Channels calculations have been performed in order to better understand the features of the optical potentials found in the OM analysis. This study searches to elucidate some aspects of the optical potential of weakly bound systems, such as the dispersion relation and the long range (attractive and absorptive) mechanisms.
 
\end{abstract}


\date{\today}
\end{frontmatter}

\maketitle

\section{Introduction}
Previous optical model (OM) analyses of elastic scattering data for many heavy-ion systems have shown a rapid and localized variation of the optical potential at energies in the vicinity of the Coulomb barrier, known as the ``threshold anomaly'' \cite{Sat91}. This effect is characterized by a rapid increase  of the surface strength of the imaginary part of the potential with increasing energy, up to an approximately constant value,  accompanied by a bell-shaped peak in the surface strength of the real part. These energy-dependent surface strengths are linked by a dispersion   relation ~\cite{Nag85}.  Physically, this variation can be understood in terms of a dynamic polarization potential arising from channel 
coupling effects \cite{Sat91,Nag85,Lil85,Kee96,Kee98}. 
As the bombarding energy increases, the absorption of flux tends to be constant and the energy dependence of the real potential is found to be mainly connected with non-local effects~\cite{Can97}. A realistic non-local model for the heavy-ion real potential developed in Refs.~\cite{Can97,Cha97,Cha98,Gal98,Cha02} has shown to be a powerful tool for studying elastic scattering. In Ref.~\cite{Cha02}, 
a simple local approximation to this non-local model was derived. This potential factorizes  in a fundamental energy-dependent term and an energy-independent double folded potential $V_{F}(r)$, as following:
\begin{equation}
\label{eq:VN}
V_{SPP}(r)=V_{F}(r)\exp{\left(-\frac{4v^{2}}{c^{2}}\right)} ,
\end{equation}
where $c$ is the speed of the light and $v$ is the relative velocity between the two nuclei. This microscopic model, called S\~ao Paulo Potential (SPP), has  been tested for a large number of systems \cite{Can97,Cha97,Cha98,Gal98,Cha02,Alv03,Alv99,Sil02,Ros02},  demonstrating its validity at low (sub-Coulomb) and intermediate energies. In particular, it has been applied to the scattering of $^{6}$He on $^{12}$C and $^{58}$Ni targets, at different energies~\cite{Alv05,Gas03}, without the need of any renormalization of the real bare potential. The use of this microscopic potential tends to reduce the ambiguities in the real part of the bare potential \cite{Sat83}.

Previous works, e.g.\ Refs.~\cite{Kee96,Tho89}, have shown that the ``threshold anomaly'' behaviour of the interaction potential involving stable heavy-ion projectiles can be explained by coupled-channels calculations employing double folded real potentials without renormalizations, provided that the physically significant couplings are explicitly included. These calculations usually employ an interior imaginary potential of Woods-Saxon (WS) form to simulate the in-going wave boundary condition for fusion, all surface absorption being provided by the channels that are explicitly included in the coupling scheme. 

Although the appearance of the threshold anomaly has been studied extensively in the scattering of light weakly bound  $^{6,7}$Li 
\cite{Kee94,Mar98,Mac99,Pak03,Pak04,Gom05,Fer07}, there are very few works done with $^{6}$He~\cite{Dip03,So05}.
The $^{6}$He nucleus has a weakly bound three-body  n-n-$\alpha$ Borromean structure and it is known to have an extended two neutron distribution. This peculiar structure affects the dynamics of the collision, making these 
processes very interesting,  as demonstrated by recent experimental campaigns. The data from these experiments show a sizable sensitivity to this underlying exotic structure~\cite{So05,Ver97,Dav00,Esc07}. Reactions induced by $^{6}$He on several targets, at energies around the Coulomb barrier, exhibit some common features, such as a remarkably large cross section for the production of $\alpha$ particles. This effect is clearly associated with the weak binding of the halo neutrons ($S_\mathrm{2n}=0.975$ MeV), that favours the dissociation of the $^{6}$He projectile in the nuclear and Coulomb fields of the target. 

Moreover, we have recently learned that simple preconceptions based on the experience of the OM on stable nuclei, such as the role of the strong absorption radius, cannot be simply extrapolated to the scattering of exotic nuclei. We have also learned about the long range reaction mechanisms which occur when $^{6}$He is scattered on a 
heavy target, such as $^{208}$Pb \cite{Esc07,San08,Kak06}. These recent results confirm that the elastic scattering induced by exotic nuclei on heavy targets can be qualitatively different from the scattering of stable nuclei. It is observed that the elastic data are sensitive to the values of the real and the imaginary potentials in different radial ranges. In general, the sensitivity to the real potential corresponds to the region of the strong absorption radius, which is $R_{sa}=12.5$~fm for the $^{6}$He+\nuc{208}{Pb} reaction \cite{Esc07}. However, it is found that the long range absorption present in the $^{6}$He scattering data can only be reproduced using large values for the diffuseness of the  imaginary part of the phenomenological potential, and the sensitivity to the imaginary potential occurs at much larger radii \cite{Kak06}. Again, we are able to assess that these data can not be well reproduced by short range potentials.

In the present OM analysis, we try to understand the distinctive features of the $^6$He optical potential, for both light and heavy targets. For this purpose, we consider the  targets $^{27}$Al and  $^{208}$Pb. In order to keep the underlying ingredients 
as fundamental as possible, trying to avoid any kind of ambiguities in the real part of the bare interaction, we adopt the double-folding potential of Eq.~(\ref{eq:VN}), without any renormalization. This bare interaction is supplemented with the analytical Coulomb dipole polarization (CDP) potential of Ref.~\cite{And94}, which describes the effect of the dipole Coulomb interaction on the elastic scattering. This CDP potential is entirely determined by the $B(E1)$ distribution and does not contain any free adjustable parameter. The remaining part of the interaction is described by a complex WS potential, whose parameters are determined by a best fit procedure of the elastic data. In order to obtain good fits of 
the $^6$He+$^{208}$Pb data, 
both the real and imaginary parts of the WS potential require a large value of the diffuseness parameter. In order to get further insight of this result, four-body Continuum-Discretized Coupled-Channels (CDCC) calculations \cite{Rod09} have been also performed. The trivially equivalent local polarization potential  derived from these calculations show  long range attractive and absorptive tails for the real and imaginary components, respectively. This result justifies the need of long-range WS potentials in the phenomenological OM analysis of this system. Under this interesting scenario, we checked the consistency of the resulting optical potential prescription with the dispersion relation.

This paper is structured as follows. In section 2, we first present a brief summary of previous $^6$He,$^6$Li+$^{27}$Al and $^6$He,$^6$Li+$^{208}$Pb OM analyses, pointing out some differences between these systems. Afterwards, we present a conventional OM analysis based on a complex optical potential, whose real bare potential is described by the microscopic SPP potential of Eq.~(\ref{eq:VN}). This double-folding potential 
is eventually supplemented with an analytical parameter-free, complex CDP potential which takes into account the effect of dipole Coulomb couplings. Unlike the  $^{6}$He+$^{27}$Al case, in order to reproduce satisfactorily the $^{6}$He+$^{208}$Pb data one needs to include real and imaginary long range components in the optical potential. 
In section 3, we present four-body CDCC calculations. The local equivalent polarization potential, derived from these calculations, is compared with the phenomenological potentials extracted in the OM analysis. Finally, in section 4 we present the summary and conclusions of this paper.

\section{Optical Model analysis}

\subsection{Previous Optical Model analyses for $^{6}$He reactions}

Within the optical model (OM) framework one must determine appropriate values for the geometry of the optical potential. Here we review some previous OM analyses of $^6$He,$^6$Li+$^{27}$Al,$^{208}$Pb systems, with the aim of pointing out some common and different features of these systems, besides defining our starting point.

It is well known that the  $^{6}$Li and $^{6}$He systems have similar structures. They are both weakly bound, with separation energies of 1.475 MeV and 0.975 MeV, respectively. The main qualitative difference between them is that the dipole Coulomb force can break up $^{6}$He into $^{4}$He+2n, but it cannot break up $^{6}$Li into $^{4}$He+$^{2}$H. The dipole Coulomb operator, in a $N = Z$ nucleus, is an isospin 1 operator. Since $^{6}$Li, $^{4}$He and $^{2}$H, have isospin 0 in their ground states, it is not possible that the dipole Coulomb force breaks up $^{6}$Li into $^{4}$He+$^{2}$H. Therefore, the mechanisms of $^{6}$Li and $^{6}$He break up in an intense electric field are very different. Breakup of $^{6}$Li is governed by the nuclear interaction while breakup of $^{6}$He is governed by both the nuclear and Coulomb couplings to the continuum~\cite{Esc07}. 

Kakuee et al.~\cite{Kak06} showed that OM potentials with parameters adjusted to fit the $^{6}$Li+$^{208}$Pb elastic scattering data are not able to reproduce the $^{6}$He+$^{208}$Pb data at 27 MeV. By contrast, Benjamim et al.~\cite{Ben07} have shown that OM calculations based on the double folding potential of Eq.~(\ref{eq:VN}) describe satisfactory  $^6$He+$^{27}$Al  and $^6$Li+$^{27}$Al elastic data at laboratory energies from $E_\mathrm{lab}$=7.0 to 13.4 MeV. 


S\'anchez-Ben{\'{i}}tez et al.~\cite{San08} and Kakuee et al.~\cite{Kak06} have performed phenomenological OM calculations for  $^{6}$He on $^{208}$Pb at $E_\mathrm{lab}$=14, 16, 18, 22, and 27 MeV, using a standard Woods-Saxon (WS) shape.
The most remarkable feature of the 
derived potentials is the large value of the imaginary diffuseness parameter required to reproduce the data ($a_i \approx 2$~fm).
This result was interpreted as an evidence of the presence of long-range reaction mechanisms. This is in contrast to the $^{6}$Li+$^{208}$Pb case, where this phenomenon has not been observed. 

\begin{table}[tb]
\begin{center}
\begin{tabular}{lccccc} 
\hline
System & $v$ (MeV) & $w$ (MeV) & $a_i$ (fm) & $\chi^2$/point & Ref.\\
\hline
$^6$He+$^{208}$Pb & 81.0 & 7.0 & 1.75 & 1.5  & \cite{Dav00} \\
$^6$He+$^{208}$Pb & 22.0 & 5.5(3) & 1.89 & 1.9& \cite{San08} \\ 
$^6$Li+$^{208}$Pb & 109.5 & 22.4 & 0.88 & - &  \cite{Coo82}\\   
\hline
\end{tabular}
\end{center}
\caption{Real and imaginary WS potential parameters used to fit $^6$He+$^{208}$Pb elastic scattering cross section at 27 MeV from Refs.~\cite{San08,Dav00} and $^6$Li+$^{208}$Pb elastic scattering from Ref. \cite{Coo82} are shown. The rest of parameters are fixed at $a_{r}=0.811$ fm and $R_{0}=R_{0i}=7.856$ fm.}
\label{tab1}
\end{table}

\subsection{Optical model analysis for \nuc{6}{He}+\nuc{208}{Pb} and \nuc{6}{He}+\nuc{27}{Al}}

 We start with a conventional OM analysis, using a complex optical potential, based on the microscopic S\~ao Paulo potential (SPP), given by Eq.~(1). We have considered 
the experimental data from  Refs.~\cite{San08,Kak06} for  \nuc{6}{He}+\nuc{208}{Pb} and Ref.~\cite{Ben07} for 
 \nuc{6}{He}+\nuc{27}{Al}.
The  energy-independent part of the bare nuclear interaction, $V_F(r)$, was calculated with a double folding procedure
using the effective nucleon-nucleon interaction of Ref.~\cite{Cha02}, and the matter densities of the  $^{6}$He and 
$^{208}$Pb, taken from \cite{Gas03} and \cite{Cha02}, respectively.  Both densities are parametrized in terms of Fermi-Dirac distributions.  For the imaginary part, we take the same geometry as the real part, with a normalization factor of  $N_{I}=0.78$. This is the model OM 1.
In the past, this prescription has been able to describe a large variety of systems in a very wide range of energies (see Ref.~\cite{Alv03} for details). 

\begin{figure}
\begin{center}
\includegraphics[width=1.0\textwidth]{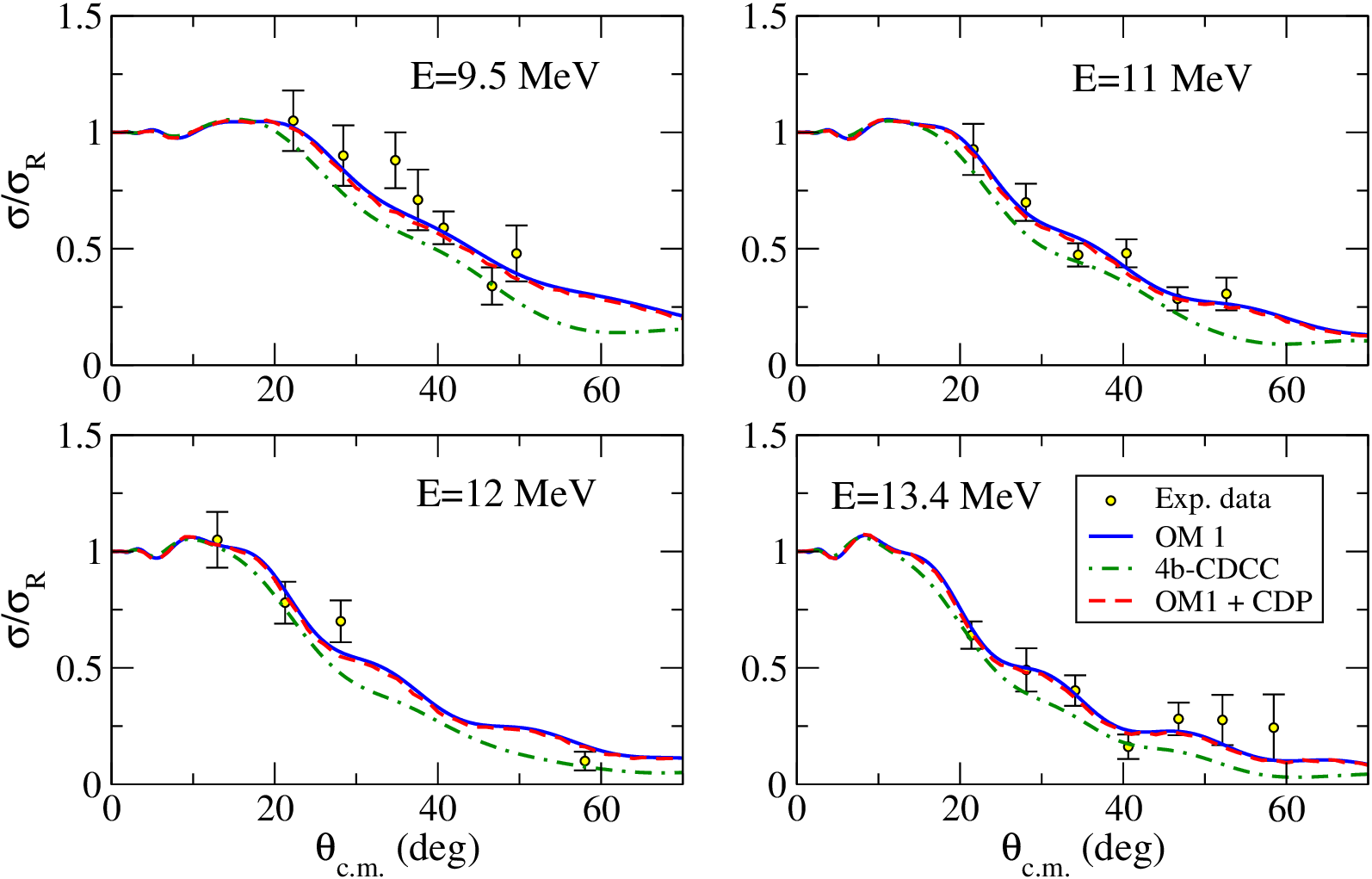}
\caption{Elastic scattering angular distribution in the center of mass frame 
for  the reaction $^{6}$He+$^{27}$Al at different bombarding energies ($E_{\rm lab}$=9.5, 11, 12, and 13.4 MeV). The circles are the experimental data from Ref.~\cite{Ben07}.
The OM  calculation, using  double-folding real and imaginary components (OM 1), are displayed by solid lines. The dot-dashed lines are the four-body CDCC calculations. By dashed lines are shown the OM calculations using the OM 1 model plus the CDP potential.}
\label{he6al_elManu}
\end{center}
\end{figure}

\begin{figure}
\begin{center}
\includegraphics[width=\textwidth]{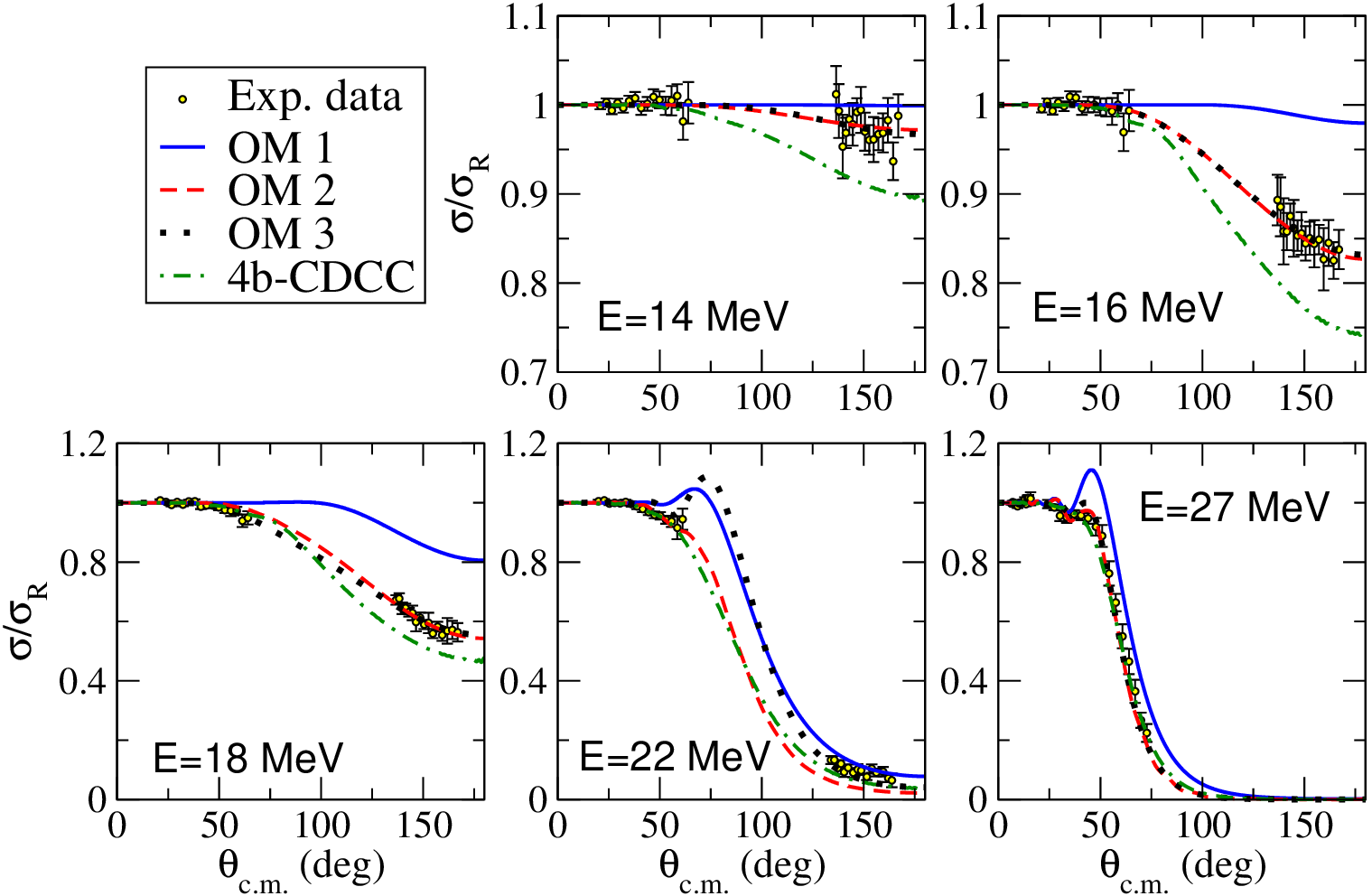}
\end{center}
\caption{Elastic scattering angular distribution in the center of mass frame 
for the reaction $^{6}$He+$^{208}$Pb at different bombarding energies ($E_{\rm lab}$=14, 16, 18, 22, and 27 MeV). The  circles are the experimental data from Refs.~\cite{Kak06,San08}.
The different OM model analyses are presented by solid (OM 1), dashed (OM 2) and dotted (OM 3) lines. The four-body CDCC calculations are shown by dot-dashed lines.}
\label{he6pb_el}

\end{figure}

In Figs.~\ref{he6al_elManu} and \ref{he6pb_el}, we compare these OM  calculations (solid lines) with the experimental data (open circles) for $^{6}$He+$^{27}$Al  and  $^{6}$He+$^{208}$Pb systems, respectively, at several bombarding energies.  It can be readily noted that these calculations reproduce very well the 
 \nuc{6}{He}+\nuc{27}{Al} data, in 
agreement with the results reported by Benjamim et al.~\cite{Ben07}, but they clearly 
fail to reproduce the $^{6}$He+$^{208}$Pb data. For the higher energies, the calculation shows a rainbow peak that is not present in 
the data. Moreover,  the elastic scattering cross section is overestimated by the calculations, mainly at backward angles, 
suggesting  that the flux removed from the elastic scattering due to the reaction channels is underestimated.

\begin{table}[tb]
\begin{center}
\begin{tabular}{ccc} 
\hline
 $E_{\rm lab}$ (MeV) & $w_L$ (MeV)  & $\chi^2$/point   \\
\hline
14 &  1.5  & 0.5  \\      
16 &  3.2  & 0.4  \\
18 &  4.1  & 1.1  \\ 
22 &  5.0  & 9.2  \\ 
27 &  5.4  & 4.2  \\ 
\hline
\end{tabular}
\caption{Results of the OM analysis for the reaction $^6$He+$^{208}$Pb using an energy-independent geometry. For all the fits, the only free parameter is the depth of the imaginary potential, $w_L$. The radius and the diffuseness of the imaginary potential were fixed to $R_{i0}=7.856$~fm and $a_{i}=1.9$~fm, respectively. There is no renormalization in the bare potential.}
\label{tab2}
\end{center}
\end{table}

\begin{table}[tb]
\begin{center}
\begin{tabular}{cccc} 
\hline
$E_{\rm lab}$ (MeV) & $w_L$ (MeV) & $a_i$ (fm) & $\chi^2$/point  \\
\hline
14 &  1.0  & 2.0 & 0.5   \\
16 &  2.70 & 2.0 & 0.4   \\
18 &  1.63 & 2.6 & 0.8   \\ 
22 &  0.30 & 2.5 & 3.2  \\ 
27 &  10.28 & 1.6 & 3.9  \\ 
\hline
\end{tabular}
\caption{Results of the OM analysis for $^6$He+$^{208}$Pb at energies from 14 to 27 MeV, using an energy-dependent geometry for the phenomenological imaginary potential. The only free parameters are the depth and diffuseness of the imaginary potential. The radius of the imaginary potential was kept fixed to 
$R_{i0}=7.856$ fm. There is no renormalization in the bare potential.}
\label{tab3}
\end{center}
\end{table}

To improve the agreement with the data, we have performed a second analysis, in which we keep the double-folding SPP for the real 
part of the nuclear interaction, but we describe the imaginary part by means of a standard Woods-Saxon (WS) shape, with adjustable 
parameters.  We also include an interior imaginary potential with a WS shape to simulate the in-going boundary condition for fusion. So, in this new analysis the projectile-target potential is parametrized as:
\begin{equation}
\label{VsppW}
U(r)= V_\mathrm{SPP}(r) + i W_S(r) + i W_L(r) ,
\end{equation}
with 
\begin{equation}\label{EQWS}
W_{S}(r)=-\frac{w_S}{1+\exp{\left(\frac{r-R_s}{a_s}\right)}} \quad \quad
W_{L}(r)=-\frac{w_L}{1+\exp{\left(\frac{r-R_{i0}}{a_i}\right)}} \, .
\end{equation}
 
The parameters of the interior WS potential ($W_{S}(r)$) are kept fixed at all energies to the values: $w_S=50$ MeV, $R_s=3.87$ fm, and $a_s=0.2$ fm.




The parameters of the external WS potential ($W_{L}(r)$) are adjusted for each 
energy to fit the elastic data. 
 With the aim of checking consistency, we assumed the same geometry of the imaginary WS potential obtained by S\'anchez-Ben{\'{i}}tez et al.~\cite{San08} (Table~\ref{tab1}) for the $^{6}$He+$^{208}$Pb system, obtained at  22 and 27 MeV, where we expect the elastic scattering to be most sensitive to the geometry of the nuclear potential. Thus, the reduced radius and the 
diffuseness of the imaginary potential were fixed to 
$R_{i0}=7.856$ fm and $a_{i}=1.9$ fm, respectively. 

With the optical potential geometry fixed, we allowed the imaginary depth ($w_L$) to vary in order to reproduce the elastic scattering data at laboratory energies of 14, 16, 18, 22, and 27 MeV. These OM fits have been performed with the routine {\sc SFRESCO}, which is part of the {\sc FRESCO} code \cite{Tho88}.  This is the model OM 2. 
The extracted values of the depths are summarized in Table~\ref{tab2} and the corresponding angular distributions are compared with the data in Fig.~\ref{he6pb_el} (dashed lines). The fits reproduce very well the experimental angular distributions, but with some limitations at 27 MeV and backward angles at 22 MeV.

As a third step, and  in order to study the sensitivity of the data to the geometry, we have allowed the diffuseness of the imaginary potential to vary, along with the depth. The diffuseness value remained around 1.9 fm (1.6 - 2.6 fm). The extracted  parameters are summarized in Table~\ref{tab3}. This is model OM 3. The new fits, shown in Fig.~\ref{he6pb_el} by dotted lines, provide a better agreement with data.  However, at 22 MeV, this calculation shows a pronounced rainbow, which does not follow the trend of the data with the energy at the measured angles. These large 
values of the imaginary diffuseness parameter are consistent with the prescription 
of  Bonaccorso and Carstoiu \cite{Bon02}. Using a semiclassical approach, and assuming that 
the imaginary part of the optical potential is well represented by an exponential form, they estimated that 
the diffuseness of the  imaginary potential is approximately given by   $a_i \approx (2 \gamma_i)^{-1}$, with  
$\gamma_i=\sqrt{2 \mu |\varepsilon_b|}$, where $\varepsilon_b =\hbar^2 \gamma_i^2/2\mu$ is the binding energy. If this 
formula is applied to $^{6}$He, one gets $a_i \approx 2$~fm, which is in very good agreement with the results of the present work.

\begin{figure}[tb]
 {\centering \resizebox*{0.85\columnwidth}{!}
 {\includegraphics[angle=0]{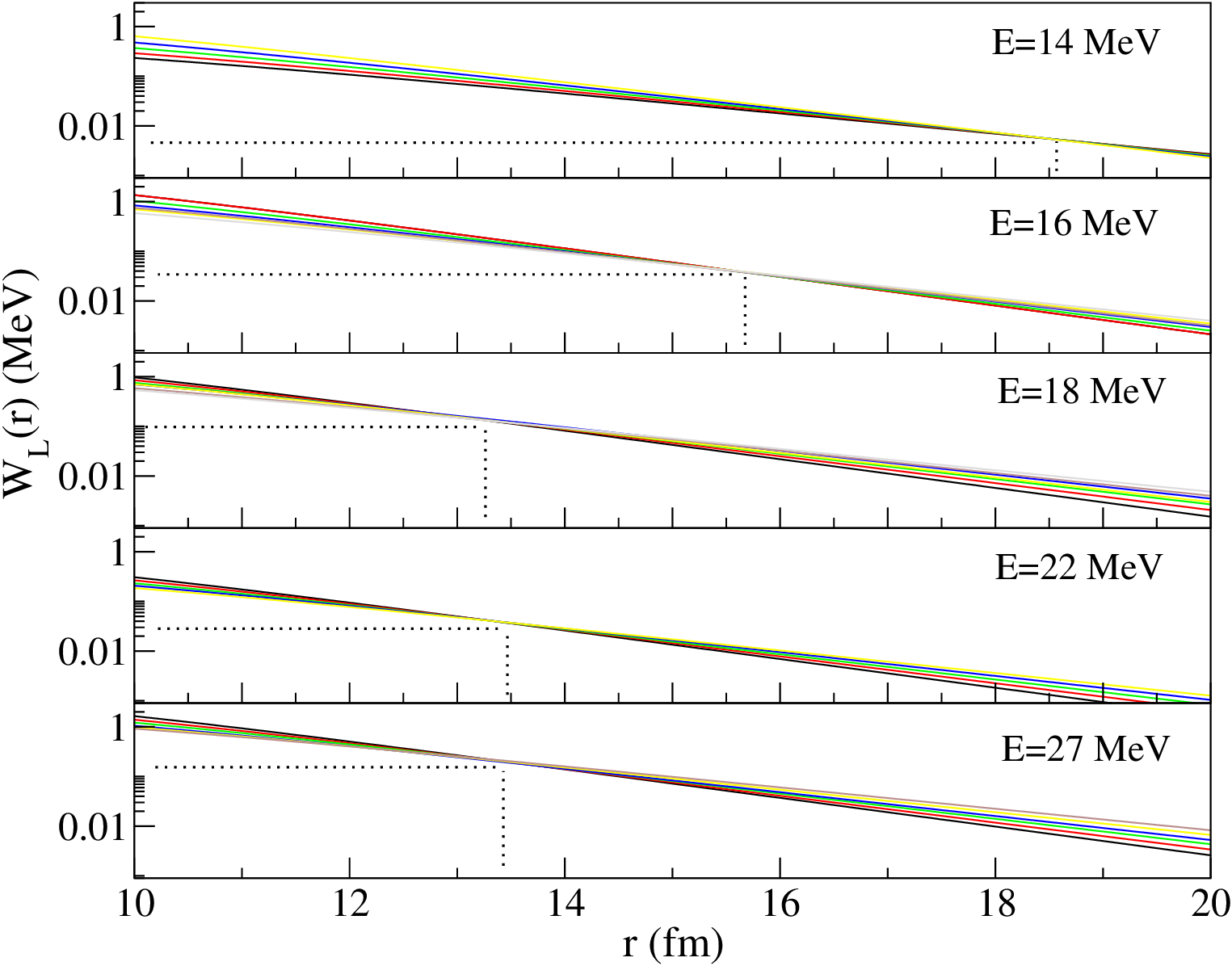}} \par}
\caption{Strengths of different imaginary optical potentials that fit the data, as a function of the interacting distance, for the 
$^{6}$He+$^{208}$Pb system. The diffuseness parameter for the imaginary part used in the calculations are in the vicinity of the $\chi^2$ minimum obtained from the best fit analysis ($a_i=1.50 - 3.0$ fm).}  
\label{rd_allgraph}
\end{figure}

The importance of this diffuse imaginary potential is better seen in Fig.~\ref{rd_allgraph}, where we plot the strength of the imaginary WS component 
as a function of the interacting distance, for different values of the diffuseness parameter $a_i$. The values of $a_i$ chosen for this plot are close to the $\chi^2$ minimum ($a_i=1.5 - 3$~fm).  Despite the well known ambiguity 
in the choice of the diffuseness parameter \cite{Sat83} we find that, in order to reproduce satisfactorily the data set, one has to use a large value of the diffuseness parameter $a_i$.  In addition, for each energy, the radius of 
sensitivity of this potential, defined at the distance at which the different potentials cross, is well beyond the strong absorption radius, meaning that the elastic  scattering will be affected by the details of this potential at large distances.

\subsection{Optical Model analysis including a Coulomb dipole polarization potential}
It is well known that the  elastic scattering of  weakly 
bound nuclei on heavy targets is strongly affected by the polarization induced by the dipole part of the Coulomb interaction. 
 In particular, it has been found that this dipole polarizability effect gives rise to a significant reduction of the elastic scattering cross sections, which is particularly important for the  collision of weakly bound nuclei on heavy targets. 
This effect has been shown 
to account for part of the long range behaviour found in phenomenological OM analyses of these systems \cite{San08,Kak06}. 

Here, we investigate how this effect varies  for two extreme cases of the charge of the target, and in which case we identify long range Coulomb polarization as a distinctive feature of the scattering of $^{6}$He, 
at energies around the barrier.
Within the OM framework, 
the effect of the dipole polarizability on the elastic observables can be included by means of a Coulomb dipole polarization (CDP) potential. 
A simple analytical 
expression for this CDP potential was derived in \cite{And94,And95}. 
The form of the polarization potential is obtained in a semiclassical framework requiring that the second
order amplitude for the dipole excitation-deexcitation process and the first order amplitude associated with the polarization potential are equal for all classical trajectories corresponding to a given scattering energy. This leads to an analytic formula for the polarization potential for a single excited state~\cite{And94}. The expression so obtained can be generalized for the case of
excitation to a continuum of breakup states~\cite{And95} giving rise to the following formula:
\begin{eqnarray}
\label{eq:Upol}
U_{pol}(r)&=&-\frac{4\pi }{9}\frac{Z_t^{2}e^{2}}{\hbar v}\frac{1}{(r-a_{o})^{2}r} \\
&\times&\int ^{\infty }_{\varepsilon _{b}}d\varepsilon \frac{dB(E1,\varepsilon )}{d\varepsilon }
\left[ g\left(\frac{r}{a_{o}}-1,\xi \right)+if\left(\frac{r}{a_{o}}-1,\xi \right) \right] , \nonumber
\end{eqnarray}
where $a_0$ is the distance of closest approach in a head-on collision, $v$ is the projectile velocity and \emph{g} and \emph{f} are analytic functions defined as
\begin{eqnarray}
f(z,\xi ) &=& 4\xi ^{2}z^{2}\exp{(-\pi \xi )}K_{2i\xi }''\left(2\xi z\right), \\
g(z,\xi ) &=& \frac{P}{\pi }\int _{-\infty }^{\infty }\frac{f(z,\xi ')}{\xi -\xi '}d\xi ',
\end{eqnarray}
and \( \xi =\frac{\varepsilon a_{o}}{\hbar v} \) is the Coulomb adiabaticity
parameter corresponding to the excitation energy \( \varepsilon  \) of the nucleus.
An important feature of this potential is that when the breakup energy \( \varepsilon _{b} \)
is large enough, the purely real adiabatic dipole potential is re-obtained. In
the opposite limit, for small breakup energies \( f\left(\frac{r}{a_{o}}-1,\xi \right)\rightarrow 1 \)
and \( g\left(\frac{r}{a_{o}}-1,\xi \right)\rightarrow 0 \), and the polarization potential
becomes purely imaginary, depending on $r$ as \( \frac{1}{(r-a_{o})^{2}r}. \)
For our analysis we take theoretical values of the $B(E1)$ distribution of $^{6}$He~\cite{Dan98,Tho00}. This determines completely the CDP potential at different energies without using free parameters.

In this section, we reanalyze the $^{6}$He+$^{27}$Al and $^{6}$He+$^{208}$Pb elastic data, including explicitly the effect of dipole polarizability by means of the CDP potential described above. In the $^{6}$He+$^{27}$Al case, the optical potential contains the double folding SPP potential, an imaginary part with the same geometry, and the CDP potential. The real and imaginary parts of the double-folding potential were renormalized by   $N_{R}=1.0$ and  $N_{I}=0.78$, respectively. The results of these calculations are represented in Fig.~\ref{he6al_elManu}  by dashed lines. As expected, we find a negligible effect of the CDP potential  in the elastic scattering angular distributions.


\begin{figure}[tb]
\begin{center}
\includegraphics[width=\textwidth]{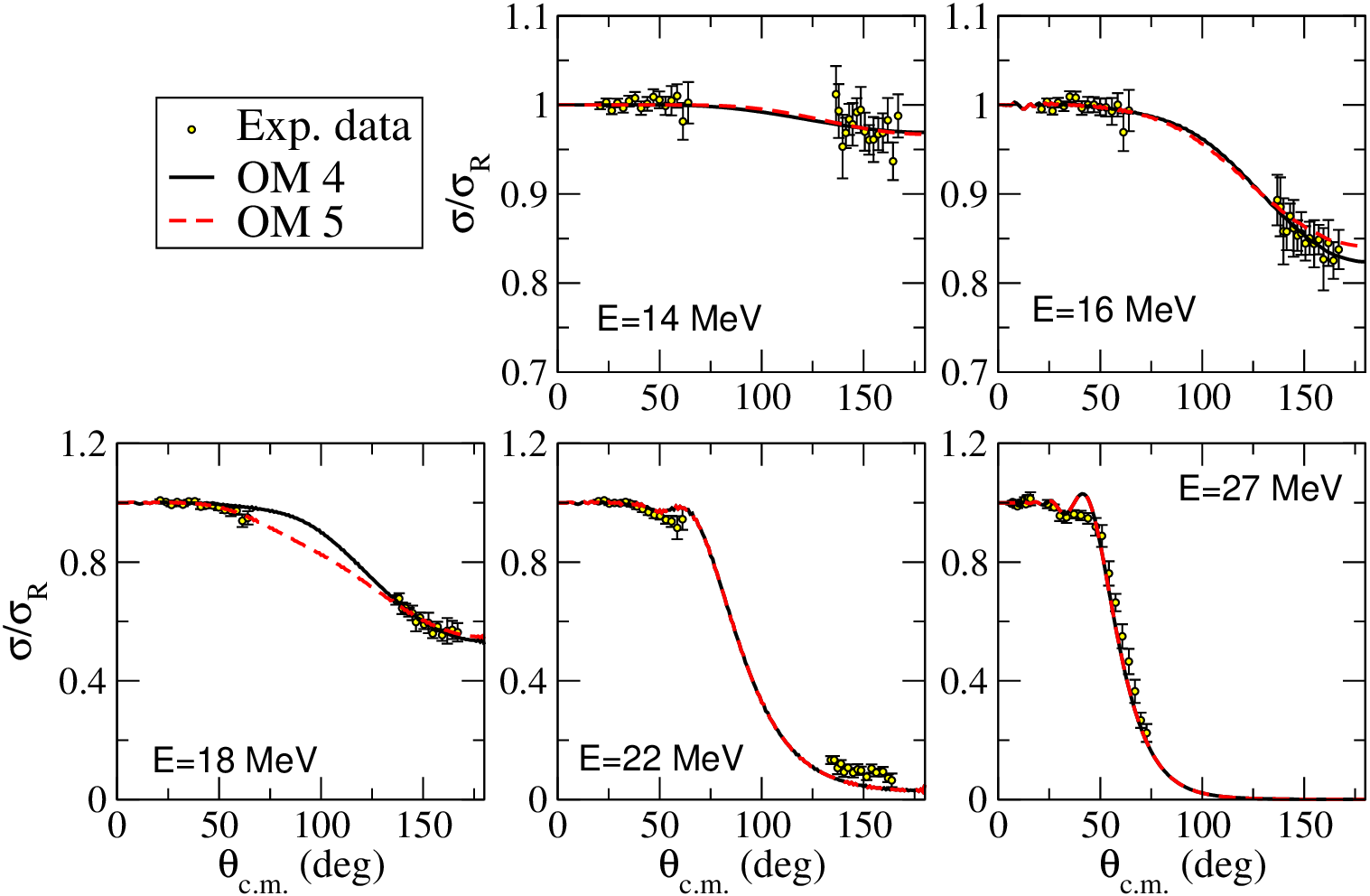}
\caption{Elastic scattering angular distribution in the center of mass frame 
for the reaction $^{6}$He+$^{208}$Pb at different bombarding energies ($E_{\rm lab}$=14, 16, 18, 22, and 27 MeV). The  circles are the experimental data from Ref.~\cite{Kak06,San08}.
The different OM model analyses, using the prescription of Eq.~(\ref{VsppUpolW}), are presented by solid (OM 4) and  dashed (OM 5) lines.}
\label{spp_dpp}
\end{center}
\end{figure}
\begin{table}[tb]
\begin{center}
\begin{tabular}{cccc} 
\hline
$E_{\rm lab}$ (MeV) & $w_L$ (MeV) &   $\chi^2$/point  \\
\hline
14 &  9.8 & 0.6  \\
16 & 37.0 & 0.4  \\ 
18 & 26.6 & 1.0  \\ 
22 & 20.0 & 9.9  \\    
27 & 24.5 & 4.1  \\ 
 \hline
\end{tabular}
\caption{Best-fit optical potential parameters for the system $^6$He+$^{208}$Pb, including explicitly the CDP potential and using energy-independent geometries for the optical potential. For all the fits, the only free parameter is the depth of the imaginary potential. The radius and diffuseness parameter of the imaginary potential were fixed to $R_{i0}=7.856$ fm and $a_i=1.1$~fm, respectively.}
\label{tab4}
\end{center}
\end{table}

For the $^{6}$He+$^{208}$Pb reaction, the optical potential contains the real double-folding SPP potential (with $N_R=1$),  the CDP potential, and  WS imaginary components, i.e.:
\begin{equation}
\label{VsppUpolW}
U(r)= V_\mathrm{SPP}(r) + U_\mathrm{CDP}(r) + i W_S(r)  + i W_L(r) .
\end{equation}
The long range component $W_L(r)$ will account for 
the effect of other peripheral reaction channels not included in the CDP potential. As before, we have divided our procedure into two steps.  In the first search, we allowed the imaginary geometry to vary for each energy. In a second search, we try to keep fixed the optical potential geometry with the energy, searching for the $\chi^{2}$ minimum. The results of these calculations are shown in Fig.~\ref{spp_dpp}. The solid lines (OM 4) correspond to 
the calculations keeping fixed the geometry of the WS potential, whereas the dashed lines (OM 5) are the calculations allowing the diffuseness parameter $a_i$  to vary. In both cases, we observe a similar agreement with data. The depths for the optical potentials extracted with the energy-independent geometry are listed in Table \ref{tab4}. Comparing these results with those of the previous subsection, we see that the inclusion of the CDP potential produces a similar quality of data fits, but 
leads to a reduction of the imaginary diffuseness from 1.9~fm to about 1.1~fm.  We can conclude that the long range tail found in the OM analysis of the  $^{6}$He+$^{208}$Pb data  is partially due to the effect of dipole Coulomb breakup.
Although the overall agreement with the data is good, it seems that some ingredient is missing, since we still have some limitations to reproduce the data at energies above the Coulomb barrier (Fig.~\ref{spp_dpp} at 22 and 27 MeV).

   

\subsection{Optical Model analysis with a long range real potential}
\begin{figure}[tb]
\begin{center}
\includegraphics[width=\textwidth]{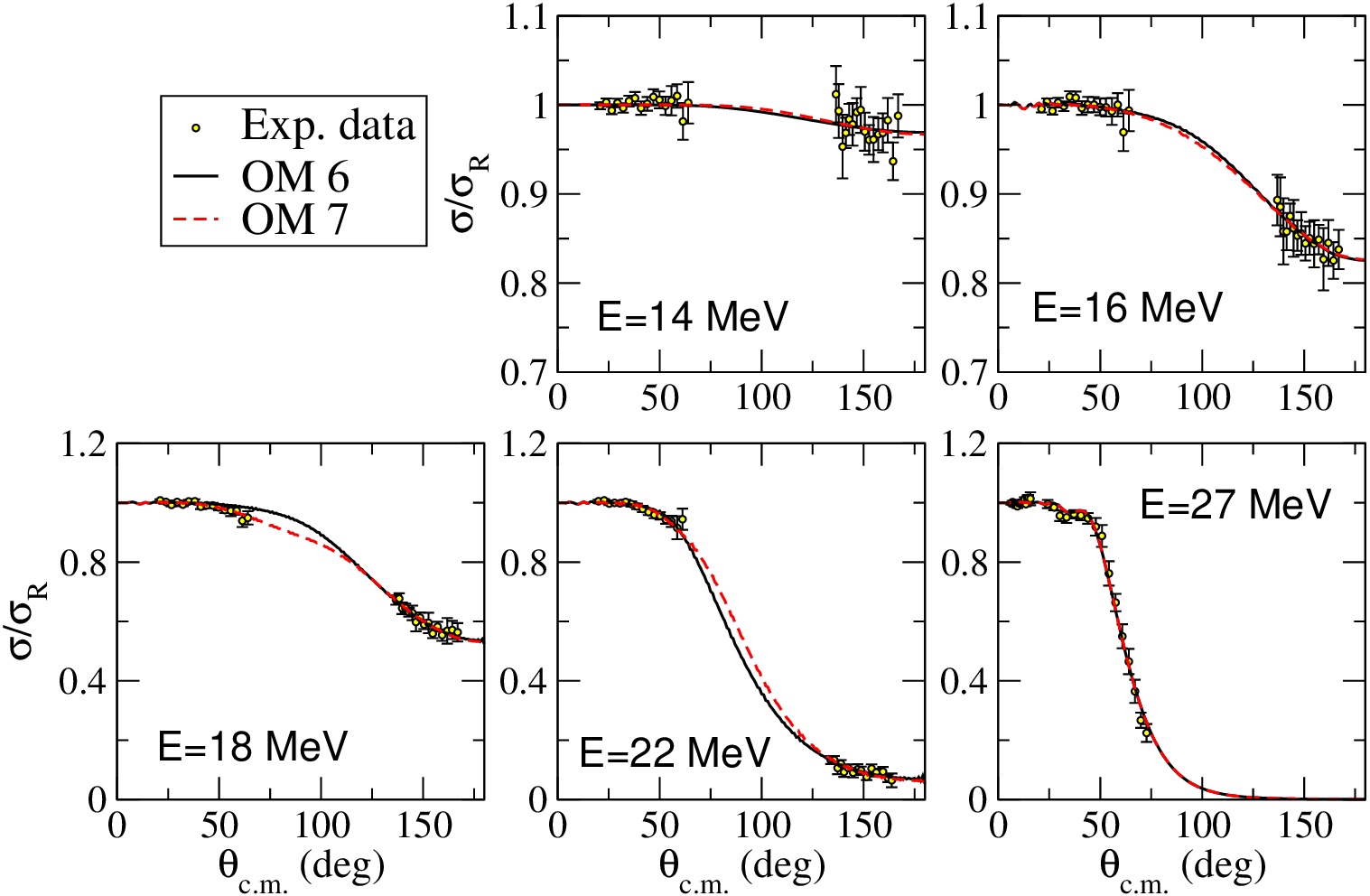}
\caption{Elastic scattering angular distribution in the center of mass frame 
for the reaction $^{6}$He+$^{208}$Pb at different bombarding energies ($E_{\rm lab}$=14, 16, 18, 22, and 27 MeV). The  circles are the experimental data from Ref.~\cite{Kak06,San08}.
The different OM model analyses are presented by solid (OM 6) and  dashed (OM 7) lines.}  
\label{sppdppvl}
\end{center}
\end{figure}

 Given the impossibility to reproduce the data above the barrier, we propose in this subsection a more general prescription in which, besides the phenomenological imaginary part, we introduce also a phenomenological real part of WS shape.
In these new serial of calculations, we keep the complex analytical CDP potential to include Coulomb breakup effects. Consequently, this analysis was performed using the following parametrization for the  optical potential:
\begin{equation}
U(r)= V_\mathrm{SPP}(r) + U_\mathrm{CDP}(r) + i W_S(r) + V_L(r) + i W_L(r) ,
\end{equation}
where $W_S(r)$ and $W_L(r)$ describe, as before, the short range and long range components of the imaginary nuclear 
polarization potential, and $V_L(r)$ is a real phenomenological  polarization potential,
\begin{equation}
 V_L(r)= -\frac{v_L}{1+\exp\left(\frac{r-R_0}{a_r}\right)} . 
\end{equation}
 As in our previous calculations, we do not renormalize the bare potential.
  The strengths $v_L$ and $w_L$, as well as the diffuseness parameters of the WS potentials ($a_r$ and $a_i$) have been allowed to vary (OM 7). Afterwards, we searched the values of $a_r$ and $a_i$ that best fit the angular distributions at all energies, and we fixed them (OM 6). The best data fits are obtained using the values $a_r=1.1$~fm  and $a_i=1.0$~fm. The extracted parameters are summarized in Table~\ref{tab6} and the corresponding angular distributions are presented in Fig.~\ref{sppdppvl}, OM 6 by solid lines and OM 7 by dashed lines. It can be seen that these calculations reproduce very well the data in the whole angular and energy range, with no significant differences between the two calculations. This analysis suggests that, besides the inelastic couplings 
produced by the dipole Coulomb force, there are other long range mechanisms that give rise to a diffuse tail in the 
phenomenological optical potential. 


%

\begin{figure}[tb]
\begin{center}
\includegraphics[angle=0,width=0.9\textwidth]{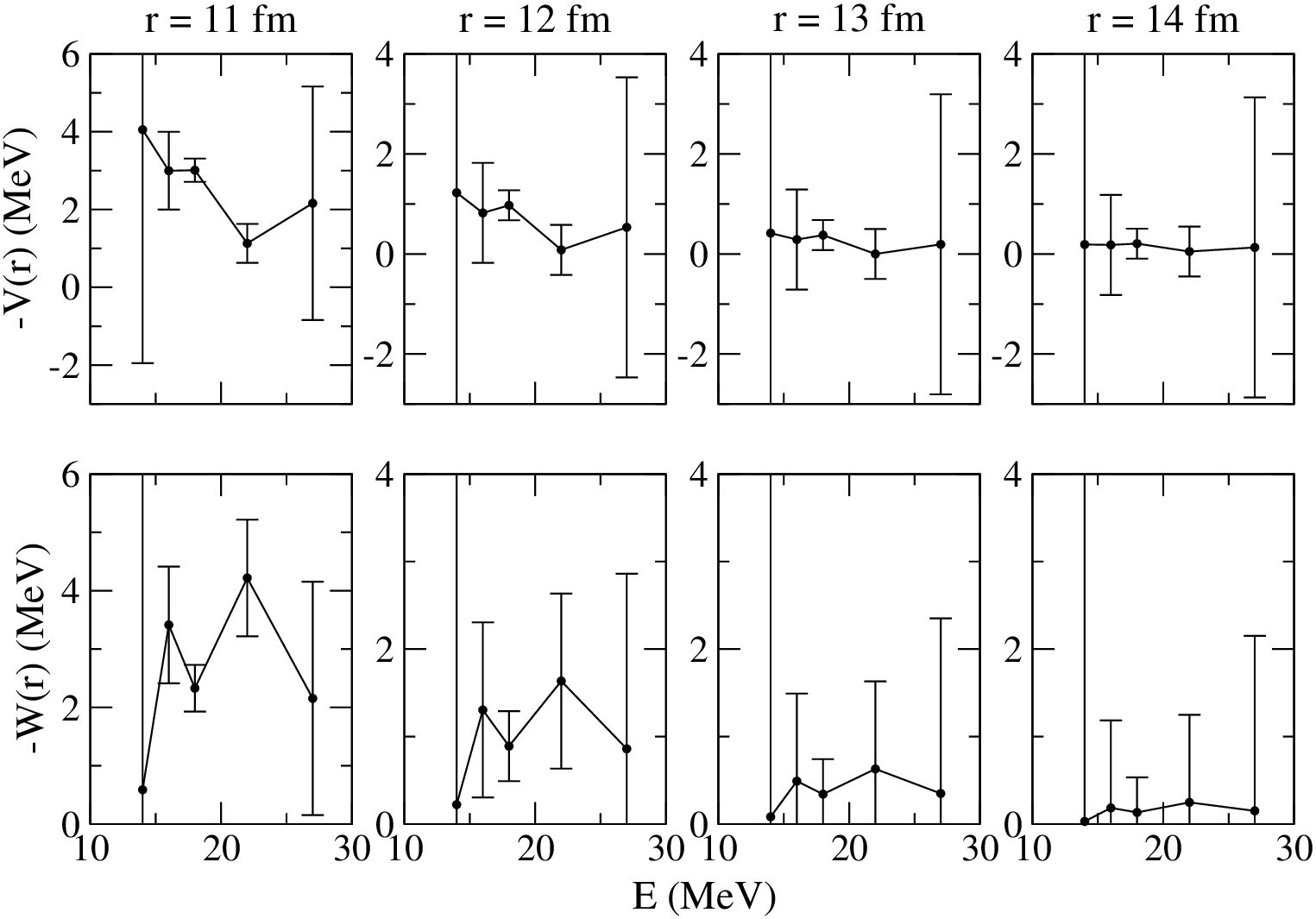}
\end{center}
\caption{Real and imaginary strengths ($V$, $W$) of the optical potential  as a function of the bombarding energy, for the system $^{6}$He+$^{208}$Pb. The strengths are evaluated at the distance indicated by the labels. The real potential strength is the sum of the double folding potential Eq.~(\ref{eq:VN}), the real part of the CDP potential and the 
phenomenological component $V_L(r)$. The strength of the imaginary potential  is the sum of the WS potentials of Eq.~(\ref{EQWS}) and the imaginary part of the dynamic polarization potential.}  
\label{rdVL}
\end{figure}

These calculations show a complex scenario from where we could expect important changes in the dispersion relation of this system as compared to normal systems.
 In Fig.~\ref{rdVL}, we plot the strength of the energy-independent geometry optical potential (as a total sum of the real and imaginary OM components), as a function of the collision energy for the $^{6}$He+$^{208}$Pb system, and for different 
values of the projectile-target distance. The results seem to show certain correlation in the variation of the real and imaginary parts, almost independent on the interacting distance, which is qualitatively consistent with dispersion relations \cite{Nag85}. However, the  errors bars limit our conclusions. 
These errors bars correspond to a variation on the complex long range potential around the $\chi^{2}_{min}$ that results on an increase of the total $\chi^{2}$ by an amount $\chi^{2}_{min}/N$. This analysis claim for better precision experiments in order to infer about the existence of a threshold anomaly in reactions involving the weakly bound $^{6}$He nucleus.

 The variation of the real and imaginary potentials with the energy has been studied also in terms of their volume integrals.  For this purpose, in Table \ref{tab6} we include also the values of the real and imaginary volume integrals ($J_v$ and 
$J_w$) per interacting nucleon pair, defined as:
\begin{equation}
\label{eq:jv}
J_{v,w}=\frac{4 \pi}{A_p A_t}\int_{0}^{\infty} U_{v,w}(r) r^2 dr 
\end{equation}
where $A_p$, $A_t$ are the projectile and target masses.  $U_v$ and $U_w$ are, respectively, the total real part (excluding the monopole Coulomb contribution) and the total imaginary part (excluding the 
interior Woods-Saxon potential) of the optical potential. For the real part, the volume integral remains roughly constant, slightly decreasing with increasing incident energy. This energy dependence is 
consistent with existing parametrizations (see eg.~\cite{Arn79,Nad02,Nad03}) and is a consequence of Passatore's \cite{Pas67,Pas75} application of Feshbach dispersion relation. Within the 
energy interval considered in the present work, the energy variation of $J_v$ is small, remaining around the value 360 MeV~fm$^3$, which is consistent with the value found 
by Mohr \cite{Moh00} in the analysis of the \nuc{6}{He}+\nuc{209}{Bi} reaction at energies around the 
Coulomb barrier. By contrast, the values of the imaginary volume integrals show a more pronounced variation with the bombarding energy. According to the dispersion relations \cite{Atz96},  the decrease of $J_v$  should be accompanied by an increase of $J_w$. In the present case, there is not a clear trend in the energy dependence of the extracted values of $J_w$. Nevertheless, one has to keep in mind 
that this energy dependence comes mainly from the Woods-Saxon potentials $V_L$ and $W_L$ and, as we 
have shown above, the strengths extracted for these potentials have large error bars.

The  large errors bars shown in Fig.~\ref{rdVL} for the real and imaginary strengths of the phenomenological potentials 
might lead to the conclusion that the features found by these potentials could be just a consequence 
of the ambiguities of the real and imaginary parts. However, in the next section, we show that these 
features, namely, the long range behaviour of the real and imaginary components, arise from physical 
effects due to specific reaction mechanisms present in the collision.

\begin{table}[tb]
\begin{center}
\begin{tabular}{ccccccc} 
\hline
$E_{\rm lab}$ (MeV) & $v_{L}$ (MeV) & $w_L$ (MeV) &   $J_v$(MeV fm$^3$) & $J_w$(MeV fm$^3$) & $\chi^2$/point    \\
\hline
14 &  20.01 & 14.23 & 365 &  27 & 0.6 \\ 
16 &  0.01 & 80.12 & 364  &  152 & 0.4  \\  
18 & -2.77 & 54.67 & 363  & 103  & 1.0  \\  
22 & -36.84 & 97.97 & 362 &  185 & 0.7 \\  
27 & -19.45 & 47.96 & 360  &  92 &2.9 \\
\hline
\end{tabular}
\caption{Best-fit parameters for the reaction $^6$He+$^{208}$Pb, including explicitly the CDP potential and
the real and imaginary WS potentials $V_L(r)$ and $W_L(r)$.  For all the fits, we considered an energy-independent geometry, 
with $R_0=R_{i0}=7.856$ fm, $a_{r}=1.1$ fm, and $a_{i}=1.0$ fm,  which provide the best overall fit. For each energy, the free parameters are the depths $v_L$ and $w_L$. The columns $J_v$ and $J_w$ refer to the real and imaginary volume integrals, as described in the text.}
\label{tab6}
\end{center}
\end{table}


\section{Microscopic analysis within a few-body approach}
\subsection{Continuum-Discretized Coupled-Channels calculations}
Theoretically, the treatment of the reactions induced by the Borromean nucleus
$^6$He requires a four-body formalism (three-body projectile plus a target).
For three-body problems (two-body projectile plus a target) the 
Continuum-Discretized Coupled-Channels (CDCC) framework \cite{Yah82,Aus87} has been
successfully used for many cases \cite{Rus04,Tak03}.
Recently this method has been extended to four-body problems \cite{Mat04,Mat06,Rod07,Rod08}. 
In particular it has been tested for the reaction $^6$He on $^{208}$Pb at 22 MeV \cite{Rod09}.
Consequently, in this work we use this four-body CDCC framework for performing
theoretical calculations for the reactions under study, $^6$He+$^{208}$Pb
and $^6$He+$^{27}$Al at different energies around the Coulomb barrier. 
We have used the 
binning procedure \cite{Rod09} for discretizing the three-body continuum 
of the projectile. This procedure has been developed very recently as an 
extension of the same method used traditionally in standard three-body CDCC 
calculations. For this purpose, the three-body continuum representation
uses the eigenchannel  expansion of the multi-channel S-matrix \cite{Rod09}.    

Here we use the same structure model for the three-body system 
$^6$He($\alpha$+n+n), as in Refs.~\cite{Rod09,Rod08}.
The Hamiltonian includes two-body potentials plus an effective three-body 
potential. Continuum states with angular momentum and parity $j^\pi=0^+$,$1^-$, and $2^+$ were considered. The 
wavefunctions for these states were generated using the codes {\sc FaCE} \cite{face}
 and {\sc sturmxx} \cite{sturm}. The maximum hypermomentum used was $K_\mathrm{max}=8$.  The parameters of the three-body interaction are adjusted to reproduce
the ground-state separation energy and matter radius 
(for $j=0^+$ states) and the resonance energy (for $j=1^-$ and $2^+$ states). The calculated ground state energy 
was 0.953~MeV and the root mean squared (rms) radius was 2.46~fm 
(assuming a rms radius of 1.47~fm for the $\alpha$ particle). 
Both Coulomb and nuclear potentials are included. The fragment-target interactions 
were represented by optical potentials which reproduce the elastic scattering at the appropriate energy.
The $n+^{208}$Pb and $n+^{27}$Al  potentials  were from \cite{Kon03}, the $\alpha+^{208}$Pb was from from \cite{Bar74}, 
and for $\alpha+^{27}$Al we used the code by S. Kailas \cite{kailas}, which
provides optical model parameters for
$\alpha$ particles using real potential volume integrals of Atzrott et al.~\cite{Atz96}, 
geometry systematics, and dispersion relation.

The coupled-channels equations were solved using the code 
{\sc FRESCO}~\cite{Tho88}, that reads the coupling potentials externally.
We included in the calculation 
the projectile-target interaction multipole couplings with order $Q=0,1,2$. 
In order to get convergence, the number of eigenchannels included was 4 for both systems.
However, the maximum energy value $\varepsilon_{\rm max}$, the number of 
bins $n_{\rm bin}$ for each $j^{\pi}$, the maximum total angular momentum $J_\mathrm{max}$
and the matching radius $R_m$ depended on the target and on the energy.
These values are presented in Table~\ref{CDCC_par}.

\begin{table}[tb]
\begin{tabular}{ccccccc} 
\hline
 System & $E_{\rm lab}$ (MeV) &  $\varepsilon_{\rm max}$ (MeV) & $n_{\rm bin}$ ($0^+,1^-,2^+$) & $J_\mathrm{max}$ & $R_m$ (fm) \\
\hline
$^6$He+$^{208}$Pb & 14.0 & 5.0 & (6,9,6) & 150 & 200    \\ 
$^6$He+$^{208}$Pb & 16.0 & 6.0 & (9,12,9) & 150 & 200\\ 
$^6$He+$^{208}$Pb & 18.0 & 7.0 & (9,12,9) & 150 & 200\\ 
$^6$He+$^{208}$Pb & 22.0 & 8.0 & (6,9,6) & 150 & 200 \\
$^6$He+$^{208}$Pb & 27.0 & 8.0 & (6,9,6) & 150 & 200 \\ 
$^6$He+$^{27}$Al & 9.5 & 5.0 & (6,9,6) & 30 & 80\\
$^6$He+$^{27}$Al & 11.0& 6.0 & (6,9,6) & 30 & 80\\
$^6$He+$^{27}$Al & 12.0& 7.0 & (6,9,6) & 30 & 80 \\
$^6$He+$^{27}$Al & 13.4& 8.0 & (6,9,6) & 30 & 80\\

\hline
\end{tabular}
\caption{Parameters for the four-body CDCC calculations. See text for details.}
\label{CDCC_par}
\end{table}

Figs.~\ref{he6al_elManu} and \ref{he6pb_el} show the  four-body CDCC calculations for the $^6$He+$^{27}$Al and $^6$He+$^{208}$Pb systems, at different energies around the Coulomb barrier (dash-dotted lines). 
For the 
$^6$He+$^{27}$Al case, the calculations reproduce well the trend of the data, although some underestimation is observed. 
For the $^6$He+$^{208}$Pb case, these calculations reproduce very well the forward angular region. However, as the energy decreases the calculation tends to underestimate the data at larger angles. 
For this reaction, we have used a second set of optical potentials for the fragment-target interactions. These were generated by the SPP using Eq.~(\ref{eq:VN}). The calculations with these potentials yield very similar results, although there are small differences at backward angles. However the magnitude of these differences cannot explain the underestimation of the data at these large angles.
\subsection{Local equivalent polarization potential from CDCC calculations}
\begin{figure}
\begin{center}
\includegraphics[width=0.45\textwidth]{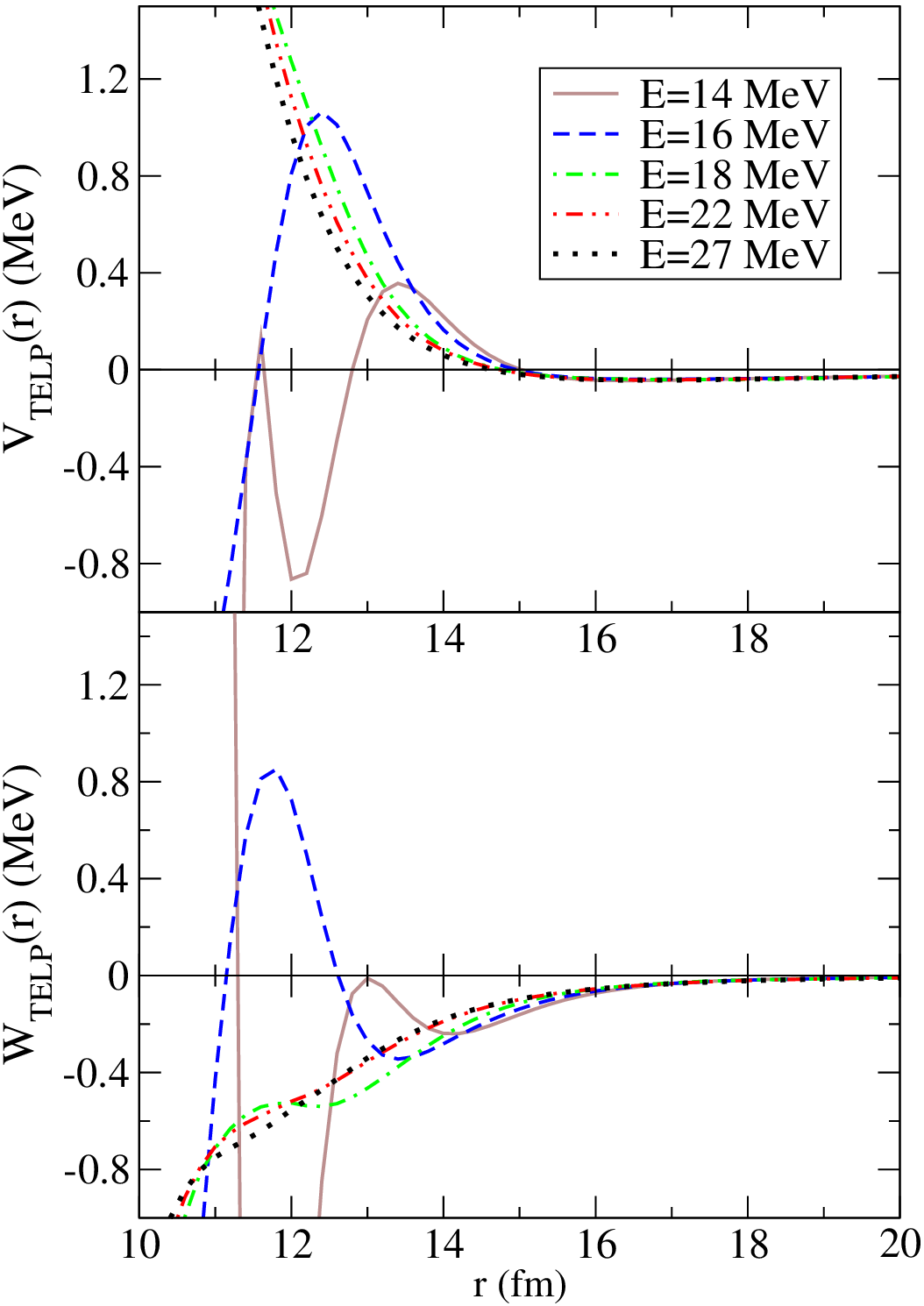}
\includegraphics[width=0.45\textwidth]{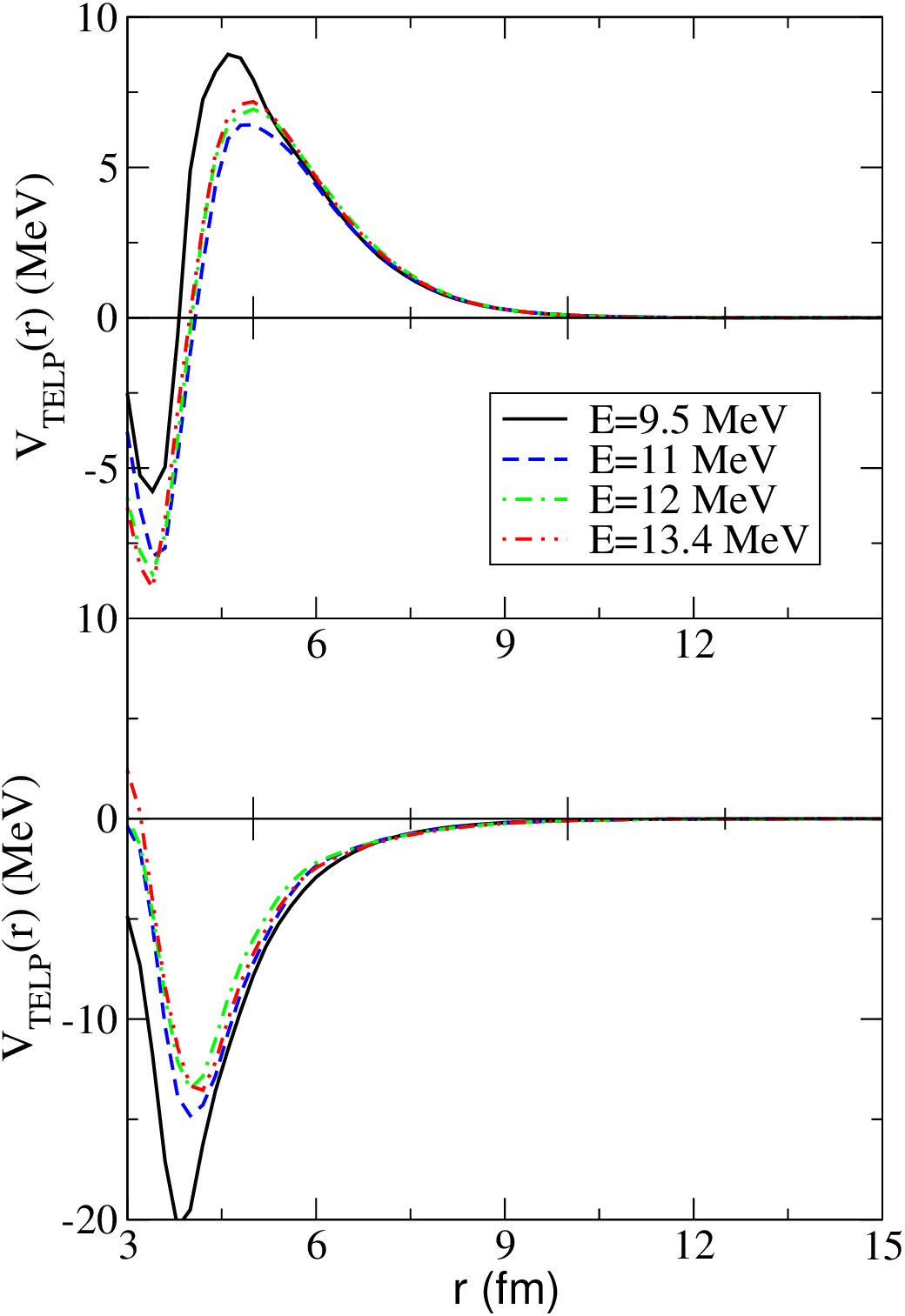}
\caption{$^{6}$He+$^{208}$Pb (left) and $^{6}$He+$^{27}$Al (right) TELP potentials extracted from the four-body CDCC calculations.}  
\label{he6pol}
\end{center}
\end{figure}

In order to link the results of this section with those found in the OM analysis, we have extracted  from the four-body CDCC calculations the so called trivially equivalent local polarization (TELP) potential \cite{Tho89}. This is a local and
 $L$-independent potential which represents the overall effect of the breakup channels on the elastic scattering.  This 
potential is constructed in such a way that the one-channel calculation performed with the potential $U_\mathrm{bare}(r)+ U_\mathrm{TELP}(r)$ gives the same elastic scattering as the full CDCC calculation. The bare potential,  $U_\mathrm{bare}(r)$ is just the sum of the 
fragment-target interactions convoluted with the ground state density of the \nuc{6}{He} nucleus. Figure \ref{he6pol} shows these polarization potentials (due to Coulomb and nuclear interaction) calculated for the different systems at different energies. 

For the $^{6}$He+$^{27}$Al system (right panel in Fig.~\ref{he6pol}), we  observe a very uniform behavior of the dynamic polarization potential (real and imaginary parts) as a function of the interacting distance. The real part of the dynamic polarization potential is repulsive (except at very short 
distances where the details of  this potential are probably not meaningful). Moreover, this potential shows a weak dependence with the incident energy. This result explains why the use of a double folding bare potential and an imaginary part with the same geometry is able to reproduce satisfactorily the data.

On  the contrary, for the $^{6}$He+$^{208}$Pb system (left panel in Fig.~\ref{he6pol}), such uniform behavior is not observed.  The real part is repulsive at distances close to the strong absorption radius and becomes attractive at larger distances. The imaginary part is mostly absorptive, although for the lowest energies it becomes emissive 
at short distances.\footnote{It has been pointed out \cite{Mac09} that this emissive imaginary part is a  consequence of representing a strongly non-local object, namely, the dynamic polarization potential arising from the coupled-channels couplings, by a simple local potential. This effect, nevertheless, does not lead to unitary 
breaking.} Both the real and imaginary parts extend to large distances, well beyond the strong absorption radius. These features are consistent with the findings of Mackintosh and Keeley \cite{Mac09} and Rusek \cite{Rus09} for the same 
reaction. Clearly, this complicated behaviour cannot be simply simulated by a renormalization of the double-folding potential.

\begin{figure}
\begin{center}
\includegraphics[width=0.45\textwidth]{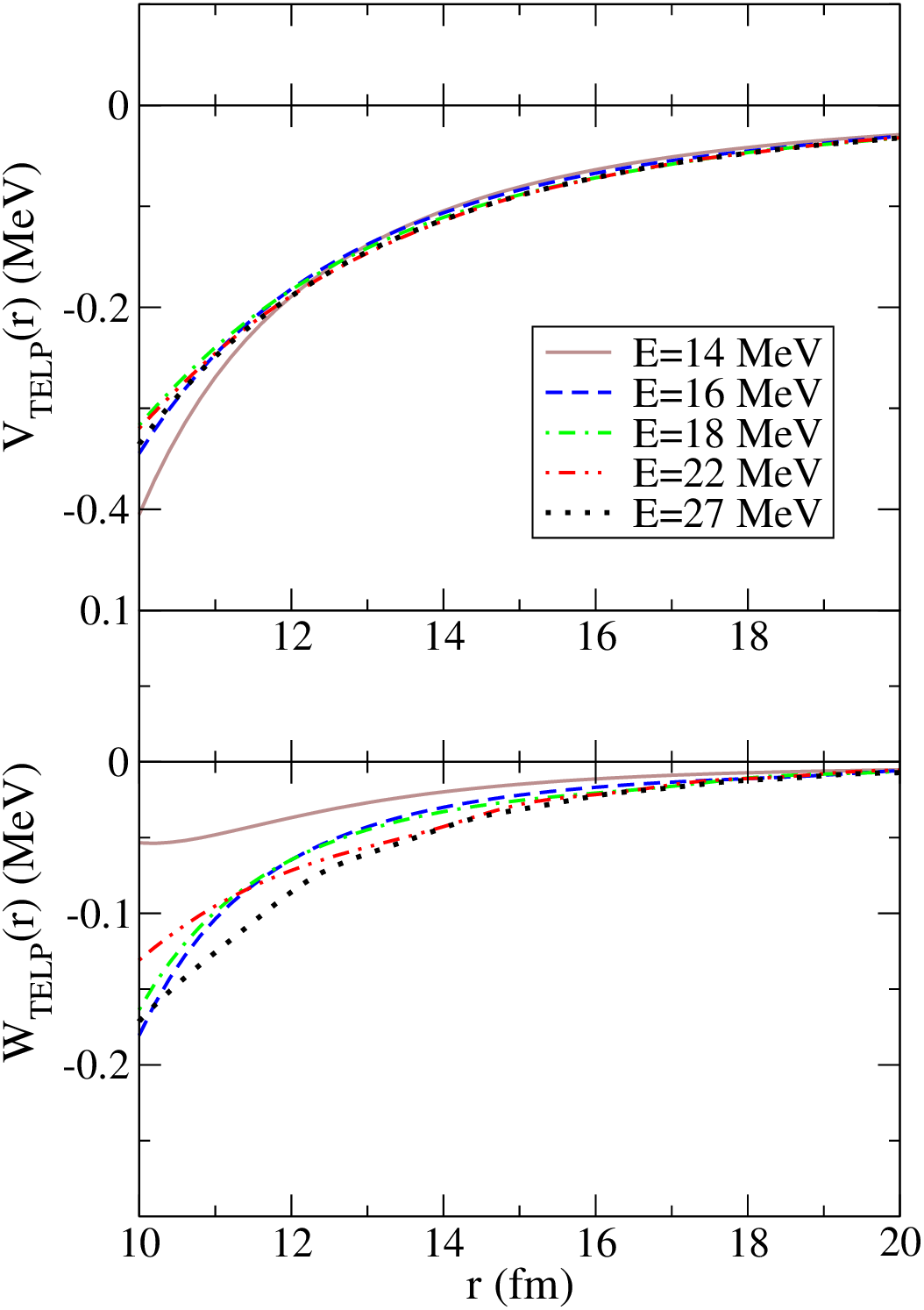}
\includegraphics[width=0.45\textwidth]{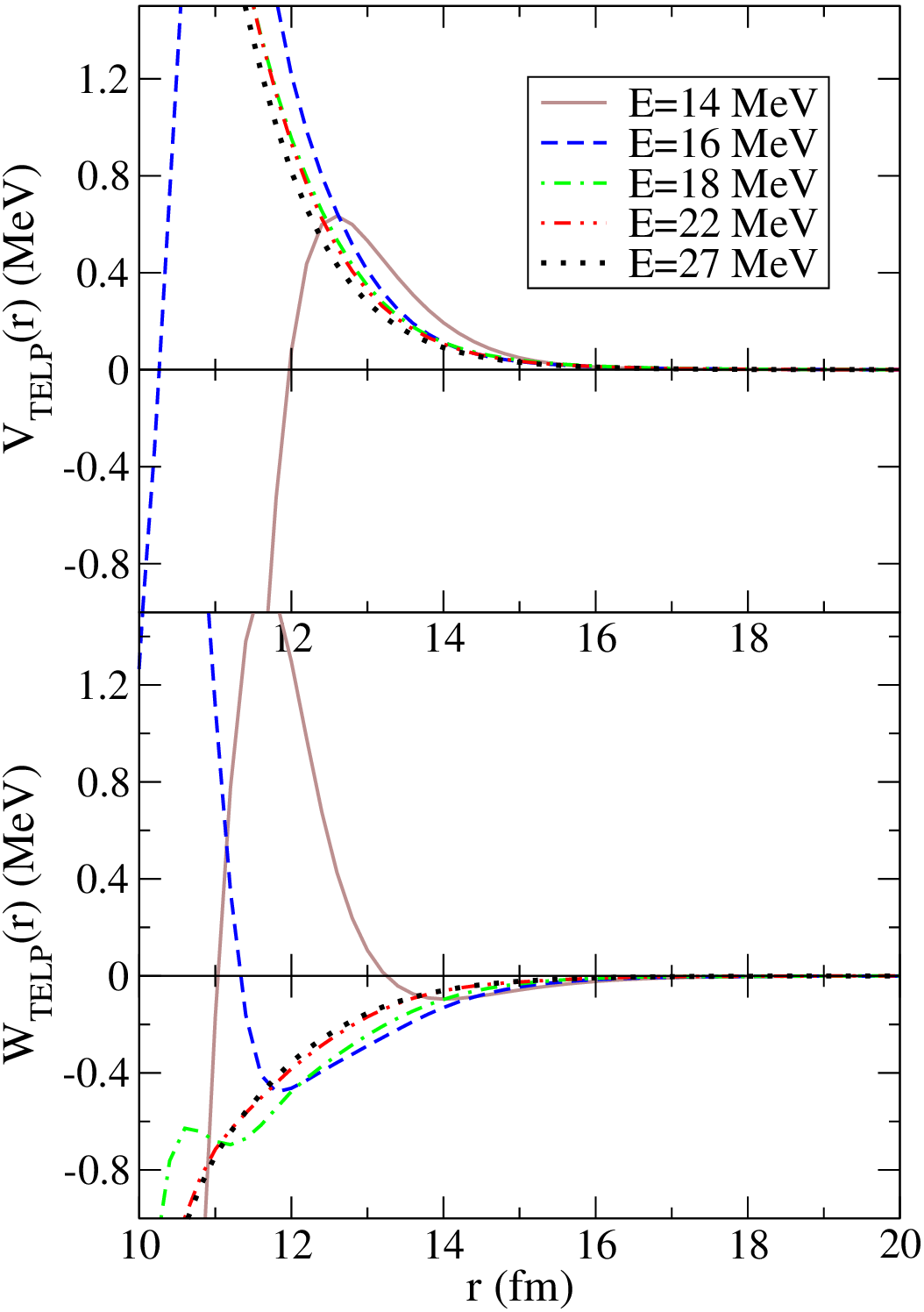}
\caption{$^{6}$He+$^{208}$Pb Coulomb (left) and nuclear (right) TELP potentials extracted from the four-body CDCC calculation.}  
\label{he6pol_C}
\end{center}
\end{figure}

To get a deeper understanding on the relationship between the features of the TELP and  the phenomenological potentials extracted in the OM analysis of the \nuc{6}{He}+\nuc{208}{Pb} reaction we  show in Fig.~\ref{he6pol_C} the separate contributions of the TELP potentials
arising from either Coulomb (left) or nuclear (right) couplings. It can be seen  that the Coulomb couplings are responsible for the long range attractive tail in the real part 
of the TELP potential. Also, these couplings produce a long range absorptive tail. Within the phenomenological OM 
analysis, this behaviour is expected to be at least partially  taken into account by means of the analytical CDP potential. Also from 
Fig.~\ref{he6pol_C} one sees that nuclear couplings are responsible for the strong repulsive part 
of the polarization potential. Although they are of shorter range than the Coulomb polarization potential, it is noticeable that both the real and imaginary components of the nuclear polarization potential extend also to distances well beyond the 
strong absorption radius. Therefore, besides long range Coulomb couplings, the $^{6}$He+$^{208}$Pb  reaction is 
characterized by long range nuclear couplings. In particular, the long range absorptive tail explains why the OM calculations using the double-folding potential for the imaginary part do not reproduce 
the data (even after inclusion of the CDP potential), and supports the need of a long range complex component in the OM potential. This long range part of the real polarization potential cannot be well accounted for by a mere renormalization of the double-folding potential.

 It is worth to note that the conclusions extracted from the TELP have to be analysed with 
caution, since this potential is just a local $L$-independent approximation of a very complicated 
non-local and $L$-dependent object. In particular, the oscillations of the TELP at short distances arise from the radial solution of the elastic equation, which appears in the denominator in the expression of the TELP.  To support the conclusions extracted from this potential, we have extracted another polarization potential fitting the elastic angular distribution  obtained with the CDCC method with a local optical potential. This potential contains 
the double-folding SPP potential and Woods-Saxon real and imaginary components, with fixed radius $R$=7.86 fm.  
The parameters obtained from this fit ($v_L$=-32.1 MeV, $a_r$=0.98~fm, $w_L$=10.2 MeV, $a_i$=1.7 fm) confirm that, in order to reproduce the effect of the continuum couplings using a single-channel optical potential, the real 
and imaginary parts need  a long-range absorptive tail.

It is illustrative to compare these results with those of previous works. The effect of the breakup channels on the optical potential has been subject of many studies in the past. Although it is not the aim of this paper to make an extensive review of these works, we cite 
some previous results which are closely related to ours. In the eighties, the Kyushu group  
studied in detail the effect of the continuum in the elastic and transfer of deuterons using the CDCC 
method \cite{Yah82,Aus87}. By extracting the polarization potential from the CDCC calculations, they 
concluded that the breakup induces a surface complex  polarization potential with a real repulsive part and an imaginary  absorptive part. 
This is consistent with our results for the light target. However, for the lead target, 
we have a significant long-range attractive component and the imaginary part is of much longer range than that found in the cited work. As 
we have already pointed out, this is a consequence of the strong Coulomb couplings. These couplings where omitted in these pioneering calculations 
by the Kyushu work and, in addition, the effect is expected to be much smaller for the deuteron case due to its larger binding energy.

Similar conclusions where achieved in the comprehensive work of Sakuragi \cite{Sak87} for $^6$Li scattering. He found that the breakup 
channels produce a strong repulsive term and an absorptive part. The values of these potentials in the strong absorption radius 
were consistent with the renormalization required to reproduce the elastic data using double-folding potentials. His conclusions can 
not be readily extrapolated to the present case because (i) only nuclear breakup was included and (ii) in any case the effect of dipole 
couplings are very much suppressed in $^6$Li with respect to $^6$He. Nevertheless, 
his results are consistent with our calculations when Coulomb breakup is switched off. 

Finally, we cite the work of Matsumoto \etal ~\cite{Mat06} where they study the elastic scattering of \nuc{6}{He} on \nuc{209}{Bi} at Coulomb barrier energies within a pseudo-state version of the four-body CDCC 
method. They find the real part of the equivalent local polarization potential, extracted from their CDCC calculations, is 
 repulsive at short distances and becomes attractive for distances beyond $\approx$15~fm, whereas the imaginary part is absorptive and of long range. These results are in total agreement with ours, as expected given the similitude between both reactions.



\section{Summary and conclusions}
Detailed optical model (OM) analyses of the $^{6}$He on $^{27}$Al and $^{208}$Pb data, at laboratory energies around the Coulomb barrier ($E_{\rm lab}$=9.5, 11.0, 12.0, and 13.4 MeV and $E_{\rm lab}$=14, 16, 18, 22, and 27 MeV, respectively), have been performed.

In the case of the light system, $^{6}$He+$^{27}$Al, we cannot recognize breakup effects on the optical potential analysis. Thus, a conventional optical potential, with the same form factor for the real and imaginary parts, based on the double-folding potential given by Eq.~(\ref{eq:VN}), have been successfully applied to describe the data of this light weakly bound system. The inclusion of a complex Coulomb dipole polarization (CDP) potential, which takes into account the effect of the dipole 
Coulomb interaction, had no significant effect on the data fits. Four-body Continuum-Discretized Coupled-Channels (CDCC) calculations  corroborate the conclusions of the OM analysis.

In the case of the heavy system, $^{6}$He+$^{208}$Pb,  breakup effects (Coulomb and nuclear) are very relevant and play an important role in the dynamics of the reaction around the Coulomb barrier. These couplings give rise to long range attractive and absorptive components in the 
optical potential required to reproduce the data. Despite this complex scenario, we tried to develop an  optical potential,
with an energy-independent geometry, consistent with the dispersion relation and suitable for studying reaction mechanisms.

For the analysis of the $^{6}$He+$^{208}$Pb data, we adopt the microscopic real nuclear S\~ao Paulo potential given by Eq.~(\ref{eq:VN}), with its 
energy-independent geometry calculated as a double-folding of the effective nucleon-nucleon interaction \cite{Cha02} with the $^{6}$He and \nuc{208}{Pb}  matter densities. We performed calculations with and without including explicitly the analytical CDP potential. Both the double-folding and the CDP components have no adjustable parameters. The fact that, even after the inclusion of the CDP potential, one still requires relatively intense values of the optical potential at large distances, means that there are other relevant long range mechanisms, besides the dipole Coulomb polarizability. To obtain a satisfactory agreement with the data above the barrier, a repulsive long range Woods-Saxon (WS) potential (with real and imaginary parts) had also to be included in the OM analysis. In this approach, the optimal values for the diffuseness parameters were $a_r=1.1$ fm and $a_i=1.0$ fm. These values are to be compared, respectively, 
with the diffuseness value of the nuclear densities of the double-folding potential, which is $a_r=0.56$ fm, using a Fermi-Dirac representation, and with the mean imaginary diffuseness obtained to fit the data set, without using the CDP potential, $a_i=1.9$ fm. In particular, this reduction of the imaginary diffuseness when including the CDP potential confirms that Coulomb breakup  accounts for a considerable part of the long range behavior found in the OM analysis.

The existence of sizable long range effects in the $^{6}$He+$^{208}$Pb interaction, suggests the presence 
of reaction mechanisms that remove flux from the elastic channel at distances well beyond the strong absorption radius ($R_{sa}=12.5$ fm) and even well below the Coulomb barrier. These results can easily be understood since, as it has been discussed in this work and in previous ones (see, e.g.\ Ref.~\cite{Esc07}), in the case of $^{6}$He+$^{208}$Pb reaction, the breakup cross section has its maximum at 18 MeV and decreases for energies around (below and above) this value, but it remains very large for these energies. It evidences the presence of a strong dynamic polarization potential, which has shown to be consistent with the trivially equivalent local polarization (TELP) potential  derived from four-body CDCC calculations (Figs.~\ref{he6pol} and \ref{he6pol_C}). For projectile-target separations around the strong absorption radius, the real part of the TELP potential is very repulsive (Fig.~\ref{he6pol}), while it becomes attractive at large distances (Figs.~\ref{he6pol} and \ref{he6pol_C}). The repulsive part has been identified as coming from nuclear couplings, whereas the attractive part arises from Coulomb couplings. The presence of the repulsive part in the polarization potential 
justifies the inclusion of the real WS potential in the optical model analysis. Both nuclear and Coulomb couplings 
produce also a long range absorptive component in the TELP potential, which again justifies the inclusion of the imaginary WS component in the optical model calculations.  Moreover, the complicate behaviour of the TELP potential as a function of the distance, as well as its strong energy dependence, suggest
that this complicated behaviour cannot be simply accounted for by a renormalization of the double folding potential. 

Therefore, our analysis suggests a new and complex scenario for the optical potential involving the scattering of the weakly bound $^{6}$He nucleus on heavy targets. The attractive and absorptive effects, produced by the couplings to the continuum breakup states, tend to produce significant changes in the strength of the real and imaginary parts of the optical potential with the energy, at the different interacting distances. Consequently, any OM analysis involving this nucleus as well as the application of the dispersion relation to the study of threshold anomaly, must be made with caution. With the aim of assessing the consistency with the dispersion relation, depending mainly on the dynamic polarization potential effects, we have studied the energy dependence of the real and imaginary parts of the potential evaluated at different values 
of the interacting distance. 
 The results, presented in Fig.~\ref{rdVL}, show a certain correlation in the variation of the real and imaginary parts with the energy, almost independent on the interacting distance, which is consistent with the dispersion relation \cite{Nag85}.  However, our conclusions are limited by the errors bars and so,  in order to allow us to infer about the dispersion relation and threshold anomaly in the $^{6}$He reactions, more accurate data would be required.


For the future, couplings between the transfer/breakup channels and the elastic channel must be incorporated beyond the first order, thus performing a coupled-reaction channel calculation. This calculation could allow an assessment on whether the explicit inclusion of these channels can account for the remaining part of the long range absorption effects.

%
%
\ack This work has been supported by the Spanish Ministerio de Ciencia e Innovaci\'on
under project FPA2006-13807-C02-01, the local government of Junta de Andaluc\'{\i}a under the excellence project P07-FQM-02894 and the Spanish Consolider-Ingenio 2010 Programme CPAN (CSD2007-00042). 
We are grateful to M.V. Andr\'es for providing us the Coulomb polarization potentials.


\bibliographystyle{elsart-num}
\bibliography{./BTA6He}

\begin{thebibliography}{10}
\expandafter\ifx\csname url\endcsname\relax
  \def\url#1{\texttt{#1}}\fi
\expandafter\ifx\csname urlprefix\endcsname\relax\def\urlprefix{URL }\fi

\bibitem{Sat91}
G.~R. Satchler, Phys. Rep. 199 (1991) 147--190.

\bibitem{Nag85}
M.~A. Nagarajan, C.~C. Mahaux, G.~R. Satchler, Phys. Rev. Lett. 54 (1985)
  1136--1138.

\bibitem{Lil85}
J.~S. Lilley, B.~R. Fulton, M.~A. Nagarajan, I.~J. Thompson, D.~W. Banes, Phys.
  Lett. B151 (1985) 181--184.

\bibitem{Kee96}
N.~Keeley, J.~S. Lilley, J.~A. Christley, Nucl. Phys. A603 (1996) 97--116.

\bibitem{Kee98}
{N.~Keeley~{\it et al.}}, Nucl. Phys. A628 (1998) 1--16.

\bibitem{Can97}
M.~A. C\^andido-Ribeiro, L.~C. Chamon, D.~Pereira, M.~S. Hussein, D.~Galetti,
  Phys. Rev. Lett. 78 (1997) 3270--3273.

\bibitem{Cha97}
L.~C. Chamon, D.~Pereira, M.~S. Hussein, M.~A. C\^andido-Ribeiro, D.~Galetti,
  Phys. Rev. Lett. 79 (1997) 5218--5221.

\bibitem{Cha98}
L.~C. Chamon, D.~Pereira, M.~S. Hussein, Phys. Rev. C 58 (1998) 576--578.

\bibitem{Gal98}
D.~Galetti, S.~S. Mizrahi, L.~C. Chamon, D.~Pereira, M.~S. Hussein, M.~A.
  C\^andido-Ribeiro, Phys. Rev. C 58 (1998) 1627--1633.

\bibitem{Cha02}
L.~C. Chamon, B.~V. Carlson, L.~R. Gasques, D.~Pereira, C.~D. Conti, M.~A.~G.
  Alvarez, M.~S. Hussein, M.~A. C\^andido-Ribeiro, E.~S. Rossi, Jr., C.~P.
  Silva, Phys. Rev. C 66 (2002) 014610.

\bibitem{Alv03}
M.~A.~G. Alvarez, L.~C. Chamon, M.~S. Hussein, D.~Pereira, L.~R. Gasques, E.~S.
  Rossi, Jr., C.~P. Silva, Nucl. Phys. A723 (2003) 93--103.

\bibitem{Alv99}
M.~A.~G. Alvarez, L.~C. Chamon, D.~Pereira, E.~S. Rossi, Jr., C.~P. Silva,
  L.~R. Gasques, H.~Dias, M.~O. Roos, Nucl. Phys. A656 (1999) 187--208.

\bibitem{Sil02}
{C.~P.~Silva~{\it et al.}}, Nucl. Phys. A679 (2001) 287--303.

\bibitem{Ros02}
E.~S. Rossi, Jr., D.~Pereira, L.~C. Chamon, C.~P. Silva, M.~A.~G. Alvarez,
  L.~R. Gasques, J.~Lubian, B.~V. Carlson, C.~D. Conti, Nucl. Phys. A707 (2002)
  325--342.

\bibitem{Alv05}
M.~A.~G. Alvarez, N.~Alamanos, L.~C. Chamon, M.~S. Hussein, Nucl. Phys. A753
  (2005) 83--93.

\bibitem{Gas03}
{L.~R.~Gasques~{\it et al.}}, Phys. Rev. C 67 (2003) 024602.

\bibitem{Sat83}
G.~R. Satchler, Direct Nuclear Reactions, Oxford University Press, New York,
  1983.

\bibitem{Tho89}
I.~J. {Thompson}, M.~A. {Nagarajan}, J.~S. {Lilley}, M.~J. {Smithson},
  Nucl.Phys. A505 (1989) 84.

\bibitem{Kee94}
N.~Keeley, S.~J. Bennett, N.~M. Clarke, B.~R. Fulton, G.~Tungate, P.~V. Drumm,
  M.~A. Nagarajan, J.~S. Lilley, Nucl. Phys. A571 (1994) 326--336.

\bibitem{Mar98}
I.~Martel, J.~G\'omez-Camacho, K.~Rusek, G.~Tungate, Nucl. Phys. A641 (1998)
  188--202.

\bibitem{Mac99}
{A.~M.~M.~Maciel~{\it et al.}}, Phys. Rev. C 59 (1999) 2103--2107.

\bibitem{Pak03}
{A.~Pakou~{\it et al.}}, Phys. Lett. B556 (2003) 21--26.

\bibitem{Pak04}
{A.~Pakou~{\it et al.}}, Phys. Rev. C 69 (2004) 054602.

\bibitem{Gom05}
{P.~R.~S.~Gomes~{\it et al.}}, J. Phys. G: Nucl. Part. Phys. 31 (2005)
  S1669--S1673.

\bibitem{Fer07}
{J.~O.~Fern\'andez-Niello~{\it et al.}}, Nucl. Phys. A787 (2007) 484--490.

\bibitem{Dip03}
{A.~Di~Pietro~{\it et al.}}, Europhys. Lett. 64 (2003) 309--315.

\bibitem{So05}
W.~Y. So, S.~W. Hong, B.~T. Kim, T.~Udagawa, Phys. Rev. C 72 (2005) 064602.

\bibitem{Ver97}
J.~Vervier, Nucl. Phys. A616 (1997) 97--106.

\bibitem{Dav00}
{T.~Davinson~{\it et al.}}, Nucl. Inst. and Meth. A454 (2000) 350--358.

\bibitem{Esc07}
{D.~Escrig~{\it et al.}}, Nucl. Phys. A792 (2007) 2--17.

\bibitem{San08}
{A.~M.~S\'{a}nchez-Ben{\'{i}}tez~{\it et al.}}, Nucl. Phys. A803 (2008) 30--45.

\bibitem{Kak06}
{O.~R.~Kakuee~{\it et al.}}, Nucl. Phys. A765 (2006) 294--306.

\bibitem{And94}
M.~V. Andr\'{e}s, J.~G\'omez-Camacho, M.~A. Nagarajan, Nucl. Phys. A579 (1994)
  273--284.

\bibitem{Rod09}
M.~Rodr{\'{i}}guez-Gallardo, J.~M. Arias, J.~G\'omez-Camacho, A.~M. Moro, I.~J.
  Thompson, J.~A. Tostevin, Phys. Rev. C 80 (2009) 051601.

\bibitem{Ben07}
{E.~A.~Benjamim~{\it et al.}}, Phys. Lett. B647 (2007) 30--35.

\bibitem{Coo82}
J.~Cook, H.~J. Gils, H.~Rebel, Z.~Majka, H.~Klewe-Nebenius, Nucl. Phys. A388
  (1982) 173--186.

\bibitem{Tho88}
I.~J. Thompson, Comp. Phys. Rep. 7 (1988) 167--212.

\bibitem{Bon02}
A.~{Bonaccorso}, F.~{Carstoiu}, Nucl.Phys. A706 (2002) 322.

\bibitem{And95}
M.~V. Andr\'{e}s, J.~G\'omez-Camacho, M.~A. Nagarajan, Nucl. Phys. A583 (1995)
  817--820.

\bibitem{Dan98}
B.~V. Danilin, I.~J. Thompson, J.~S. Vaagen, M.~V. Zhukov, Nucl. Phys. A632
  (1998) 383--416.

\bibitem{Tho00}
I.~J. Thompson, B.~V. Danilin, V.~D. Efros, J.~S. Vaagen, J.~M. Bang, M.~V.
  Zhukov, Phys. Rev. C 61 (2000) 24318.

\bibitem{Arn79}
L.~G. {Arnold}, B.~C. {Clark}, R.~L. {Mercer}, Phys.Rev. C19 (1979) 917.

\bibitem{Nad02}
A.~{Nadasen}, S.~{Balaji}, J.~{Brace}, K.~A.~G. {Rao}, P.~G. {Roos},
  P.~{Schwandt}, J.~T. {Ndefru}, Phys.Rev. C 66 (2002) 064605.

\bibitem{Nad03}
A.~{Nadasen}, S.~{Balaji}, J.~{Brace}, K.~A.~G. {Rao}, P.~G. {Roos},
  P.~{Schwandt}, J.~T. {Ndefru}, Phys.Rev. C 68 (2003) 014613.

\bibitem{Pas67}
G.~Passatore, Nuclear Physics A 95 (1967) 694.

\bibitem{Pas75}
G.~Passatore, Nuclear Physics A 248 (1975) 509.

\bibitem{Moh00}
P.~Mohr, Phys. Rev. C 62 (2000) 061601.

\bibitem{Atz96}
U.~{Atzrott}, P.~{Mohr}, H.~{Abele}, C.~{Hillenmayer}, G.~{Staudt}, Phys.Rev.
  C53 (1996) 1336.

\bibitem{Yah82}
M.~{Yahiro}, M.~{Nakano}, Y.~{Iseri}, M.~{Kamimura}, Prog. Theor. Phys. 67
  (1982) 1467.

\bibitem{Aus87}
N.~Austern, Y.~Iseri, M.~Kamimura, M.~Kawai, G.~Rawitscher, M.~Yahiro, Phys.
  Rep. 154 (1987) 125--204.

\bibitem{Rus04}
K.~Rusek, N.~Alamanos, N.~Keeley, V.~Lapoux, A.~Pakou, Phys. Rev. C 70 (2004)
  014603.

\bibitem{Tak03}
M.~Takashina, S.~Takagi, Y.~Sakuragi, Y.~Iseri, Phys. Rev. C 67 (2003) 037601.

\bibitem{Mat04}
T.~{Matsumoto}, E.~{Hiyama}, K.~{Ogata}, Y.~{Iseri}, M.~{Kamimura}, S.~{Chiba},
  M.~{Yahiro}, Phys. Rev. C 70 (2004) 061601(R).

\bibitem{Mat06}
T.~{Matsumoto}, T.~{Egami}, K.~{Ogata}, Y.~{Iseri}, M.~{Kamimura}, M.~{Yahiro},
  Phys. Rev. C 73 (2006) 051602(R).

\bibitem{Rod07}
M.~Rodr{\'{i}}guez-Gallardo, J.~M. Arias, J.~G\'omez-Camacho, R.~C. Johnson,
  A.~M. Moro, I.~J. Thompson, J.~A. Tostevin, Eur. Phys. J. S.T. 150 (2007)
  51--52.

\bibitem{Rod08}
M.~Rodr{\'{i}}guez-Gallardo, J.~M. Arias, J.~G\'omez-Camacho, R.~C. Johnson,
  A.~M. Moro, I.~J. Thompson, J.~A. Tostevin, Phys. Rev. C 77 (2008) 064609.

\bibitem{face}
I.~J. Thompson, F.~M. Nunes, B.~V. Danilin, Comput. Phys. Commun. 161 (2004)
  87--107.

\bibitem{sturm}
I.~J. Thompson, Unpublished. Users manual available from the author.

\bibitem{Kon03}
A.~J. Koning, J.~P. Delaroche, Nucl. Phys. A713 (2003) 231--310.

\bibitem{Bar74}
A.~R. Barnett, J.~S. Lilley, Phys. Rev. C 9 (1974) 2010--2027.

\bibitem{kailas}
S.~Kailas, Reference Input Parameter Library (RIPL-2), available online at
  http://www-nds.iaea.org/RIPL-2/.

\bibitem{Mac09}
R.~S. {Mackintosh}, N.~{Keeley}, Phys. Rev. C 79 (2009) 014611.

\bibitem{Rus09}
K.~Rusek, Eur. Phys. J. A 41 (2009) 399--404.

\bibitem{Sak87}
Y.~{Sakuragi}, Phys. Rev. C 35 (1987) 2161.

\end{thebibliography}
\end{document}